\documentclass[acmlarge,nonacm]{acmart}
\usepackage[ruled, vlined, linesnumbered]{algorithm2e}
\usepackage{amsmath}
\usepackage{xcolor}
\definecolor{commentgreen}{RGB}{80,130,80}
\definecolor{headblue}{RGB}{31,56,100}

\SetCommentSty{mycommfont}
\SetKwComment{Comment}{$\triangleright$~}{}
\AtBeginDocument{%
  }

\setcopyright{acmlicensed}
\copyrightyear{2018}
\acmYear{2018}
\acmDOI{XXXXXXX.XXXXXXX}

\acmJournal{POMACS}
\acmVolume{37}
\acmNumber{4}
\acmArticle{111}
\acmMonth{8}

\begin{document}

\title{A Contrastive Learning Framework for Label-Efficient Gastric Cancer Detection and Intra-Tumour Progression Analysis}

\author{Aleakhya Pandeti}
\email{f20220016@bits-pilani.dubai.ac.in}
\affiliation{%
  \institution{Birla Institute of Technology and Science, Pilani}
  \city{Dubai}
  \country{UAE}
}
\author{Pranav M. Pawar}
\email{pranav@bits-pilani.dubai.ac.in}
\affiliation{%
  \institution{Birla Institute of Technology and Science, Pilani}
  \city{Dubai}
  \country{UAE}
}
\author{Mithun Mukherjee}
\email{mithun@bits-pilani.dubai.ac.in}
\affiliation{%
  \institution{Birla Institute of Technology and Science, Pilani}
  \city{Dubai}
  \country{UAE}
}
\author{Raja Muthalagu}
\email{raja.m@bits-pilani.dubai.ac.in}
\affiliation{%
  \institution{Birla Institute of Technology and Science, Pilani}
  \city{Dubai}
  \country{UAE}
}








\renewcommand{\shortauthors}{Aleakhya et al.}

\begin{abstract}
Gastric Cancer is among the leading causes of cancer-related mortality worldwide. The complexity and variability of histopathological images pose significant challenges for both pathologists and computer-based diagnostics, creating a growing need for machine learning approaches capable of capturing meaningful representations from limited labelled data. Self-supervised Contrastive Learning (CL) has demonstrated promising results in natural image analysis; however, its application to gastric cancer histopathology remains underexplored.
This paper proposes an integrated CL framework built on a ResNet-50 encoder trained under the NT-Xent objective, without reliance on ImageNet pre-training, to learn discriminative and staining-invariant representations from H\&E-stained tissue patches. The learned embeddings are transferred to a supervised downstream classifier for binary tumour/normal (TUM/NOR) detection, achieving a test accuracy of 83.13\%, an AUC-ROC of 0.9087, and an Average Precision of 0.9063 - competitive with recent supervised and weakly-supervised approaches on comparable histopathology benchmarks. Beyond binary diagnosis, the framework applies K-Means clustering to tumour embeddings to identify four geometrically separable pseudo-stage clusters, confirmed by a Random Forest classifier achieving 97\% accuracy and a mean Dice coefficient of 0.97, demonstrating the capacity of CL representations to encode progression-relevant morphological variation without stage-level annotations. Grad-CAM explainability is integrated throughout to ensure model transparency. The complete system is deployed via a Streamlit clinical interface. Compared to recent state-of-the-art methods such as Zhang et al. (2025, 94.14\% accuracy) and Su et al. (2022, 86.36\% accuracy), the proposed framework operates with substantially fewer labelled samples while additionally providing unsupervised progression stratification and interpretable outputs, addressing a gap not covered by existing approaches.

\end{abstract}



\received{20 February 2007}
\received[revised]{12 March 2009}
\received[accepted]{5 June 2009}

\maketitle

\section{Introduction}

\begin{figure}[htbp]
  \centering
  \includegraphics[width=5 in]{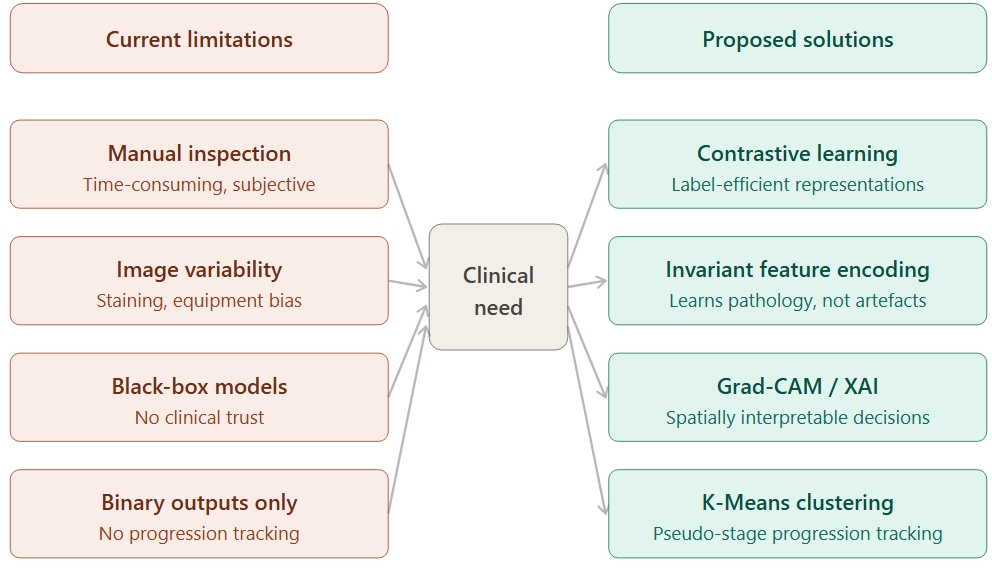}
  \caption{Diagnostic Gap and Research Motivation}
  \label{fig: mti}
\end{figure}

Gastric cancer is among the most prevalent malignancies worldwide that is consistently ranking among the leading causes of cancer-related mortality \cite{b1}. The primary diagnostic modality remains histopathology microscopic examination of tissue biopsies which constitutes the standard for cancer grading, staging, and progression assessment \cite{b2}. However, manual inspection by pathologists is inherently time-consuming, subject to variability from person to person, and sensitive to cognitive fatigue, particularly due to increasing clinical workloads and the growing complexity of tissue specimens \cite{b3}.
These limitations call for an urgent need for automated, reliable, and interpretable diagnostic tools that can support pathologists in high-throughput clinical settings. Recent advances in deep learning have demonstrated impressive capabilities in medical image analysis tasks including cancer detection, classification, and segmentation \cite{b4}. Despite this progress, two fundamental challenges persist: (i) the requirement for large-scale annotated datasets, which are expensive and labour-intensive to produce, especially in histopathology, where expert-level annotations are essential \cite{b5}; and (ii) the opaque, black-box nature of deep learning models, which undermines clinical trust and limits their utility \cite{b6}. To address these challenges, this research employs contrastive learning (CL) and explainable artificial intelligence (XAI). Figure ~\ref{fig: mti} visually represents the background and clinical motivation.

\subsection{Contrastive Learning for Label-efficient Representation}

Traditional supervised learning methods require thousands of labelled samples to achieve clinically acceptable accuracy. In gastric cancer histopathology, the cost of obtaining high-quality annotations and tracking progression stages is demands highly specialised expertise. Contrastive Learning (CL), a self-supervised learning technique, addresses this bottleneck by learning discriminative feature representations directly from unlabelled data \cite{b7}. Operating through comparison of image pairs, CL brings together representations of semantically similar samples in the latent feature space while simultaneously pushing apart representations of dissimilar samples \cite{b8}.
In the context of gastric cancer, CL-derived representations hold the potential to not only enhance diagnostic accuracy but also to model disease progression trends. By encoding histopathological images into well-separated feature subspaces, the framework can facilitate an initial stratification of cancer stages, thereby accelerating subsequent clinical verification by expert pathologists.

\subsection{Explainable AI for Clinical Trust and Transparency}
Beyond predictive performance, clinical deployment of AI systems requires transparency in decision-making. Explainable AI (XAI) techniques provide mechanisms to interpret learned features and highlight tissue regions most influential in driving model predictions. Methods such as Gradient-weighted Class Activation Mapping (Grad-CAM) \cite{b9} and SHAP (SHapley Additive exPlanations) \cite{b10} are widely applied in medical imaging contexts to generate spatially-grounded explanations.
In gastric cancer diagnosis, XAI integration serves three critical functions: (i) building clinical trust by providing interpretable justifications for AI-generated predictions; (ii) enabling pathologists to cross-validate AI findings against domain expertise; and (iii) improving model reliability by ensuring learned representations correspond to biologically meaningful features rather than misleading artifacts. When CL is coupled with XAI, the resulting framework provides not only accurate but also interpretable and clinically actionable insights, directly addressing the black-box limitation of conventional deep learning models.

\subsection{Problem Statement and Research Gap}
This paper addresses several critical gaps in existing approaches to gastric cancer histopathology image analysis:
\begin{itemize}
\item \textbf{Disease Progression Modeling}: Contemporary models predominantly provide binary or categorical cancer classifications and are not designed to capture continuous progression trajectories. Current deep learning systems lack the capacity to model stage-level gradations in gastric cancer, resulting in a critical gap between diagnostic AI output and the clinical staging process. CL-based feature encoding enables images at different stages to be placed in geometrically distinct regions of the latent space, facilitating an initial automated stratification that expedites expert verification.
\item\textbf{High Intra-class Variability}: Histopathology images of gastric cancer exhibit significant variability attributable to differences in tissue preparation, staining protocols, and imaging equipment. Such variability adversely affects the generalisation capacity of trained supervised models. CL enables models to learn staining-invariant and acquisition-invariant feature representations that better capture underlying pathological structure.
\item\textbf{Model Interpretability}: The black-box nature of standard deep learning models remains a principal barrier to clinical adoption. Integration of XAI methods within the proposed framework directly addresses this limitation, promoting transparency and enabling alignment of model decisions.
\end{itemize}

\subsection{Study Design and Key Components }
This research proposes an integrated framework for Gastric Cancer Diagnosis and potential progression tracking. The principal contributions of this work are as follows:
\begin{itemize}
  \item \textbf{Label-Efficient CL Encoder:} A ResNet-50 encoder trained under NT-Xent without ImageNet pre-training learns staining-invariant representations from stained patches, reducing reliance on costly expert annotations compared to fully supervised approaches such as Su et al.

  \item \textbf{Binary Diagnosis with Competitive Performance:} The downstream classifier achieves 83.13\% accuracy, AUC-ROC of 0.9087, and AP of 0.9063 on a held-out test set, competitive with supervised methods operating on comparable data scales, while requiring significantly fewer 
  labeled samples.

  \item \textbf{Unsupervised Progression Stratification:} K-Means clustering of tumour embeddings (k=4, determined via Elbow Method) yields four separable pseudo-stage clusters confirmed by a Random Forest classifier at 97\% accuracy and mean Dice of 0.97 which is a capability absent from all surveyed prior work on gastric cancer histopathology.

  \item \textbf{Integrated Explainability:} Grad-CAM saliency maps 
  consistently localise biologically plausible tissue structures, providing pathologist-facing transparency not present in comparable classification-only frameworks.

  \item \textbf{Clinical Deployment:} All components are integrated into a Streamlit interface providing real-time classification, cluster staging, and heatmap overlay, directly supporting the clinical workflow.
\end{itemize}

Through comprehensive analysis of Gastric Cancer Medical imaging, the research aims to provide valuable insights that contribute to a better understanding of diagnosis metrics, progression trends, and enhance decision-making for pathologists and clinicians to implement.

\subsection{Overview}
The remainder of this paper is organised as follows. Section 2 presents a comprehensive literature review of deep learning approaches for gastric cancer medical imaging, surveying existing methodologies, datasets, and identified limitations. Section 3 describes the proposed methodology, including the CL framework architecture, data processing pipeline, and XAI integration strategy. Section 4 details experimental setup and evaluation protocols. Section 5 presents and discusses results, and Section 6 concludes with directions for future research.

\section{Related Work}

This section overviews the previous work for medical imaging and disease detection, along with applications of Contrastive Learning as summarized in Table ~\ref{tab:related_work}.

\begin{table*}[]
\caption{Summary of Related Work in Gastric Cancer and Medical Imaging}
\label{tab:related_work}
\centering
\small
\begin{tabular}{|p{1.5 cm}|p{5 cm}|p{1.7cm}|p{5.2cm}|p{3cm}|}
\hline
\textbf{Reference} & \textbf{Dataset} & \textbf{Algorithm} & \textbf{Methodology} & \textbf{Evaluation Metrics} \\
\hline
Farhat et al. (2020) & Public datasets with 70\% of cases from the LIDC and ChestXray14, the most frequently used datasets in pulmonary images. & CNNs (AlexNet, VGG, ResNet, GANs) & Survey of deep learning techniques across pulmonary imaging tasks including classification, segmentation, and detection & Dice = 0.95, Accuracy up to 94\%, Sensitivity 80\% \\
\hline
Akcay et al. (2020) & The dataset consisted of 75 gastric cancer cases treated between 2012 and 2019 and with a median follow-up of 23 months &  LR, MLP, XGBoost, RF, SVC, GNB & ML-based survival and recurrence prediction with feature selection and data balancing & Accuracy = 81\%, AUC = 0.89 \\
\hline
Nguyen et al. (2020) & StructSeg dataset for organ-at-risk segmentation with (CHAOS)  and RSNA Intracranial Hemorrhage CT scan dataset & ResNet34, U-Net & Self-supervised spatial awareness pretext task followed by downstream segmentation and classification & Dice = 91.24\%, LogLoss = 0.063 \\
\hline
Pannala et al. (2020) & Datasets included 1330 images for polyp detection, 8641 labeled images with 120,000 WCE images & CNNs & Review of CNN-based GI endoscopy image analysis for detection and classification & Sensitivity = 99.8\%, Accuracy up to 94\% \\
\hline
Ben Hamida et al. (2021) & AiCOLO for colon cancer with comparisons on publicly available datasets like CRC-5000, NCT-CRC-HE-100k, and WARWICK & ResNet, U-Net, SegNet & Patch-level classification and pixel-wise segmentation of histopathology WSIs & Accuracy = 96.98\% \\
\hline
Wang \& Liu (2021) & The study collected 1340 pathological slice images of gastric cancer & DeepLabv3+ & Multi-scale semantic segmentation for gastric cancer pathological images & Dice = 91.66\%, Accuracy = 95.76\% \\
\hline
Celik et al. (2021) & Barrett's esophagus (BE) and polyps - white-light imaging (WLI) as source data and narrow-band imaging (NBI) as target data & EndoUDA (VAE + U-Net) & Unsupervised domain adaptation for modality-independent endoscopic segmentation & Dice = 0.854, IoU = 0.733 \\
\hline
Gunesli et al. (2021) & A large multi-site dataset consisting of 1.2 million image tiles from 21 different centers & FedDropoutAvg & Federated learning with randomized parameter and client dropout for histopathology & F1 = 0.91, AUC = 0.95 \\
\hline
Moon et al. (2022) & MIMIC-CXR, Open-I, and VQA-RAD & MedViLL (BERT-based) & Vision-language pretraining using masked modeling and image-report matching & AUROC = 0.98, F1 = 0.839 \\
\hline
Su et al. (2022) & WSIs from 467 gastric cancer patients were divided into a training cohort (348 WSIs), a testing cohort (88 WSIs) & ResNet-18 & Interpretable DL for tumor differentiation and MSI recognition using WSIs & F1 = 0.8977, Accuracy = 86.36\% \\
\hline
Auzine et al. (2023) & Kvasir-V2 dataset & Ensemble CNN + SHAP & Ensemble learning with explainable AI for GI cancer classification & Accuracy = 93.17\%, F1 = 97\% \\
\hline
He et al. (2024) &  A total of 207 patients were enrolled from a multicenter study  & Transformer-based DL & Attention-based multi-lesion and time-series CT analysis for survival prediction & AUC = 0.894 \\
\hline
Chen et al. (2024) & Serum samples were collected from patients with different degrees of gastric lesions & PCA + MLMNN & SERS spectral analysis combined with ML for gastric lesion classification & Accuracy = 97.5\%, Sensitivity > 96\% \\
\hline
Chai et al. (2024) & Fifteen public cancer datasets (six single omics and nine multi-omics) from TCGA and GEO databases used & COLCS (Contrastive Learning) & Self-supervised contrastive learning for multi-omics cancer subtyping & Avg. $-\log p = 2.172$ \\
\hline
Zhang et al. (2025) & A private dataset of 1889 patients with 2150 gastric ulcer, 1790 early gastric cancer, and 2713 advanced gastric cancer & MSGC (CL-based Multimodal) & Multimodal fusion of endoscopic images and reports using contrastive learning & Accuracy = 94.14\%, F1 = 94.26\% \\
\hline
\end{tabular}
\end{table*}

Farhat et al. \cite{b11} introduces early research carried out in Medical Imaging. Some of these methods include X-rays that are used for producing 2D-images by using electromagnetic waves that penetrate the body , CT (Computer Tomography) scans that use electromagnetic waves, and so on. The paper further discusses the application of deep learning models in medical imaging analysis. DL models, especially Convolutional Neural Networks (CNNs), are the most popular models used for classification, segmentation, detection, and feature extraction. DL model for lung segmentation in CXRs achieved 94\% accuracy. DL models are applied for detecting various diseases like cancer, pneumonia, and TB. All the applications of DL models provide insight into their application and potential in medical image diagnosis.

Akcay et al. \cite{b12}  focuses on evaluating the overall survival (OS) and recurrence pattern of as many as 75 gastric cancer cases treated with radiation therapy and chemotherapy using Machine Learning algorithms for prediction and for medical diagnosis. The research used logistic regression, support vector classification (SVC), multilayer perceptron (MLP), random forest, XGBoost, and Gaussian Naive Bayes (GNB) algorithms to predict the OS. For OS prediction, the GNB model achieved 81\% accuracy with an AUC of 0.89.

Nguyen et al. \cite{b13} introduces a novel self-supervised learning (SSL) approach that is mainly designed for medical image analysis, particularly CT and MR scans. The method for organ-to-risk segmentation achieved a Dice Score Coefficient of 91.02\% with 50\% of the training data and 91.24\% with 100\% of the training data, and for intracranial hemorrhage detection the spatial awareness method achieves a LogLoss of 0.128 with 50\% of training dat and 0.063 with 100\% of the training data, these results shows that the methods is capable of producing generalizable and stable performance.

Pannala et al. \cite{b14} explores the applications of AI in gastrointestinal endoscopy and specifically focuses on the analysis of endoscopic image. The paper uses the Medline database and evaluates three clinical situations for colorectal neoplastic disorders. The study uses a DL-based CNN system that improved the polyp detection rate from 29\% to 45\% and in the case of polyp classification, it led to a higher accuracy and faster image processing duration. The CNN-based model achieved a sensitivity of 99.8\% and 99.90\% for per-patient and per-lesion analysis for WCE images.

Ben Hamida et al. \cite{b15} explores the applications of Deep Learning methods for classifying and segmenting colon cancer regions in histopathological images. The research explores multiple models, including CNNs, ALEXNET, VGG, RESNET, DENSENET, and INCEPTION models, and UNET and SEGNET models for pixel-wise semantic segmentation for patch-level classification. The CNN models were tested on the AiCOLO colon cancer datasets and achieved an accuracy of 96.98\% with RESNET. With other datasets as well, the model achieved accuracies between 96\% and 99\%.

Wang \& Liu \cite{b16} introduces an automated gastric cancer segmentation model, which is based on a DeepLabv3+ neural network that enhances the efficiency of segmenting and recognizing cancerous regions in pathological slice images of gastric cancer. The DeepLabv3+ model achieved a specificity of 92.31\%, a sensitivity of 91.45\%, and an accuracy of 95.76\% and also achieves 91.66 for Dice Coefficient. The DeepLabv3+ also exhibited superior GPU operation by only requiring 2.42GB of video memory and a training time of 12.42 hours, which makes it the most efficient in terms of memory usage and training duration.

Celik et al.  \cite{b17} uses EndoUDA, which is a novel target-independent unsupervised domain adaptation (UDA) technique designed for modality-agnostic segmentation in gastrointestinal (GI) endoscopy imaging. The proposed architecture integrates a variational autoencoder (VAE) and a U-Net-based segmentation model while both use a shared EfficientNet-B4 encoder. The EndoUDA achieved good performance for different datasets, for example, for the BE dataset, it achieved 0.733 for IoU and 0.854 for Dice, and similarly, for the polyp datase,t it achieved 0.693 for IoU and 0.693 for Dice. This indicates the model’s potential in handling modality-agnostic approaches, which is crucial for effective GI cancer screening.

Gunesli et al. \cite{b18} introduces FedDropoutAvg, which is a federated learning approach that is designed to improve the generalizability and performance of deep learning models for histopathology image classification. The best parameters for FedDropoutAvg were found to be (Client dropout rate) cdr = 0.2 and (Federated dropout rate) fdr = 0.3, which were selected via grid search on the validation set. The method achieved a mean F1 score of 0.9102 and a mean AUC of 0.9542 on independent test sets, and on local test sets, it achieved a mean AUC of 0.9645 and a mean F1 score of 0.8482. 

Moon et al.  \cite{b19} introduces Medical Vision Language Learner (MedViLL), which is a multi-modal pretraining model that is designed for vision-language understanding, especially in domains like radiology images and associated reports. The model is pre-trained using two primary objectives, which are Masked Language Modeling and Image Report Matching. The MedViLL achieved an average AUROC of 0.980 and an average F1-score of 0.839 on the MIMIC-CXR dataset for diagnosis classification. Overall, the paper incorporates modality by training not just images but also reports.

Su et al. \cite{b20} introduces a deep learning system to provide interpretable recognition of tumor differentiation grade and microsatellite instability (MSI) in gastric cancer from whole-slide images (WSIs). The model was built on a pre-trained ResNet-18 model, and it was fine-tuned using transfer learning. The results showed that for tumor differentiation grade recognition, the model achieved F1 scores of 0.8615 for poorly differentiated adenocarcinoma and 0.8977 for well-differentiated adenocarcinoma. And for MSI status recognition, the models achieved patient-level accuracy of 86.36\% which shows the model’s effectiveness. 

Auzine et al. \cite{b21} introduces explainable applications to models used for medical image diagnosis. The paper proposed an ensemble deep learning model that combined InceptionV3, Inception-ResNetV2, and VGG16 through averaging-based concatenation. This ensemble model was trained on the Kvasir-V2 dataset, and it achieved an accuracy of 93.17\% and an F1-score of 97\%. Furthermore, to enhance the trust in the model, the Shapley Additive Explanations (SHAP) framework was incorporated to highlight the image regions, which were the most influential in the model’s predictions. The experiments showed that SHAP could successfully provide a visual justification of model outputs, thus building trust.

He et al.  \cite{b22} developed a deep learning model called Nomo-LDLM-2F, which is designed to predict the overall survival (OS) in patients with HER2-positive stage IV gastric cancer (GC) undergoing anti-HER2 targeted therapy, and the paper particularly focuses on longitudinal analysis using multi-lesion and time series computed tomography (CT) images. The model for one year achieved an AUC for Nomo-LDLM-2F at 0.894, and for the next, it achieved 0.809. The end result was a model that provided longitudinal analysis for the patients to learn changes in the disease. .

Chen et al. \cite{b23} introduces a novel machine learning based Surface Enhanced Raman Scattering (SERS) analysis platform designed for the label-free detection and identification of gastric lesions. The study achieved a fast, sensitive, and label-free serum spectral detection with high classification accuracy. The method demonstrated a classification accuracy of 97.5\%, a sensitivity exceeding 96.7\% and a specificity higher than 95.0\%. This research, therefore, lays a strong foundation for combining SERS with machine learning to enable real-time diagnosis and recognition of gastric lesions. 

Chai et al. \cite{b24} introduces a methodology to accurately identify cancer subtypes. In this paper, the model introduced is Contrastive Learning Based DNN (Deep Neural Network) for Cancer Subtyping (COLCS), which is a deep learning method that extracts crucial information from multi-omics features by using self-supervised learning for patient clustering. The model, compared to six other computational methods, achieved the highest average -lg p value of 2.172, with it outperforming IDEC (1.515) and scGAC (1.856), and also showing significant improvements over other methods (p < 0.05). 

Zhang et al. \cite{b25} introduces a Multi-Modal for Severity Grading of Gastric Cancer (MSGC) that aims to improve the accuracy of gastric cancer severity grading by integrating endoscopic images and diagnosis records. MSGC simultaneously extracts features from both endoscopic images and diagnostic texts using RestNeXt and BERT, respectively. The experimental results of this model show that MSGC achieved a recall of 93.69\%, a precision of 94.84\%, and an accuracy of 94.14\% along with a F1 score of 94.26\%. This approach thus gives a basic understanding of the application of CL in gastric cancer analysis.

Thus, the application of these methods helps improve model advancements into the field of medical imaging for disease diagnosis, especially in gastric cancer analysis, and helps experts to make decisions to aid in diagnosing the disease early. Provided that the field has had years of research as summarized in Table ~\ref{tab:related_work}, there is still scope to improve CL application in gastric cancer analysis, understanding, and application of future advanced solutions.

\section{Proposed Methodology}

\subsection{Research Motivation}

From the literature review conducted it is observed that most of the existing histopathology models for gastric cancer provide only binary or categorical outputs. They do not have the ability to compare and learn the different stages and progression trends in disease development across populations. Therefore, applying a model that can compare and learn enables systems to adapt over time to understand different stages and track them, which can be easily verified by clinicians at a much faster pace than raw data.
 Additionally, Deep learning models might face difficulties that are caused by variability that arises in capturing disease samples, which can be caused by the staining methods used on the equipment, which can lead to biases or deviations. Therefore, using methods like CL can help capture the underlying disease information instead of deviation from image variance due to similarity and dissimilarity learning.
 Simultaneously, Deep Learning models in histopathology often act as black boxes, which limits the trust and reliability of these models. Without clear explainability, the model predictions or results cannot be understood, as the decision process and the features that contribute to this decision are not known, which makes applying XAI a useful tool for transparency and understanding. Hence, this research focuses on incorporating these aspects into Gastric Cancer Disease Diagnosis.

\subsection{ Data Source: Gastric Cancer Histopathology Tissue Image Dataset
}

The dataset used in this study is the Gastric Cancer Histopathology Tissue Image Dataset from Kaggle \cite{b26}, which was published by researchers from the Harbin Medical University Cancer Hospital. It provides a comprehensive set of high-resolution images of gastric cancer tissues. For the study, only two categories of NORM and TUM were selected as they provide a clear binary separation for the diagnostic contrastive learning framework. 
Each image is a non-overlapping path size of 224x224 pixels that is extracted from hematoxylin and eosin-stained whole slides. The images are processed and imported using PyTorch - compatible structures and undergo some augmentations to build the CL encoder and help with generalization. The two categories themselves comprise of 4000 image samples. Overall, the GCHTID dataset supports the development of efficient diagnostic tools that make it a valuable asset for advancing AI-assisted gastric cancer analysis.

\subsection{Implementation Pipeline}

\begin{figure}[htbp]
  \centering
  \includegraphics[width=4 in]{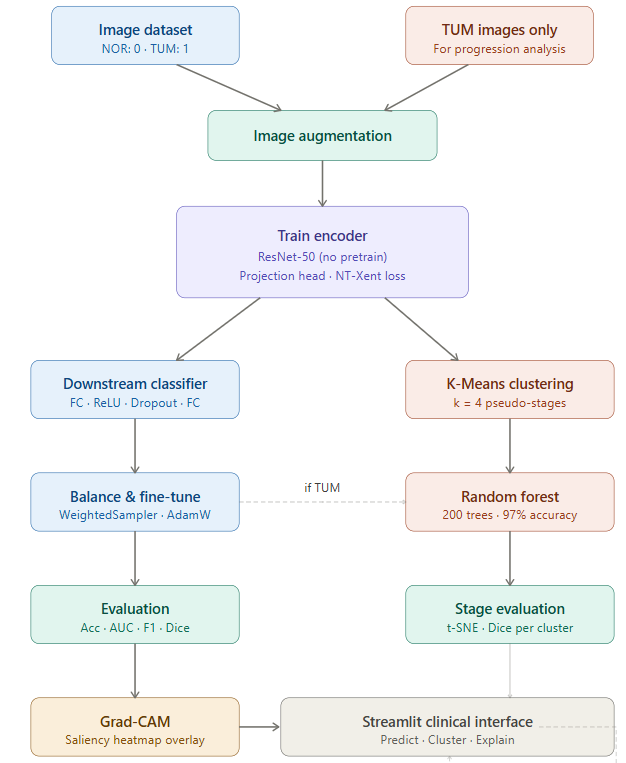}
  \caption{Implementation Architecture}
  \label{fig: arch}
\end{figure} 

\subsubsection{Data Processing and Augmentations}
The dataset was loaded using PyTorch-compatible structures, with explicit class-to-index mappings defined as NOR → 0 and TUM → 1 to ensure consistency across training and evaluation phases. During supervised fine-tuning, a WeightedRandomSampler was applied to address class imbalance by oversampling the minority class to produce 6,000 balanced training samples per epoch.
For the self-supervised contrastive learning phase, a custom ContrastiveTransform class was implemented to generate two stochastically augmented views of each input image. The augmentation pipeline comprised of: (i) RandomResizedCrop to spatial size 224×224, encouraging scale and translation invariance; (ii) RandomHorizontalFlip and RandomVerticalFlip to encode orientation invariance; (iii) RandomRotation up to 15°; (iv) ColorJitter with brightness, contrast, saturation, and hue perturbations (factor 0.5/0.5/0.5/0.1) to desensitise the encoder to staining variability; and (v) RandomGrayscale with probability 0.2. These transformations collectively encourage the encoder to extract representations invariant to non-pathological image variation while preserving diagnostically relevant tissue structure.
For supervised fine-tuning, a lighter augmentation strategy was applied (RandomHorizontalFlip, RandomVerticalFlip, RandomRotation(15°), ColorJitter(0.3/0.3/0.3/0.05)), with test images subjected only to resizing and tensor conversion to ensure unbiased evaluation.

\begin{figure}[htbp]
  \centering
  \includegraphics[width=6 in]{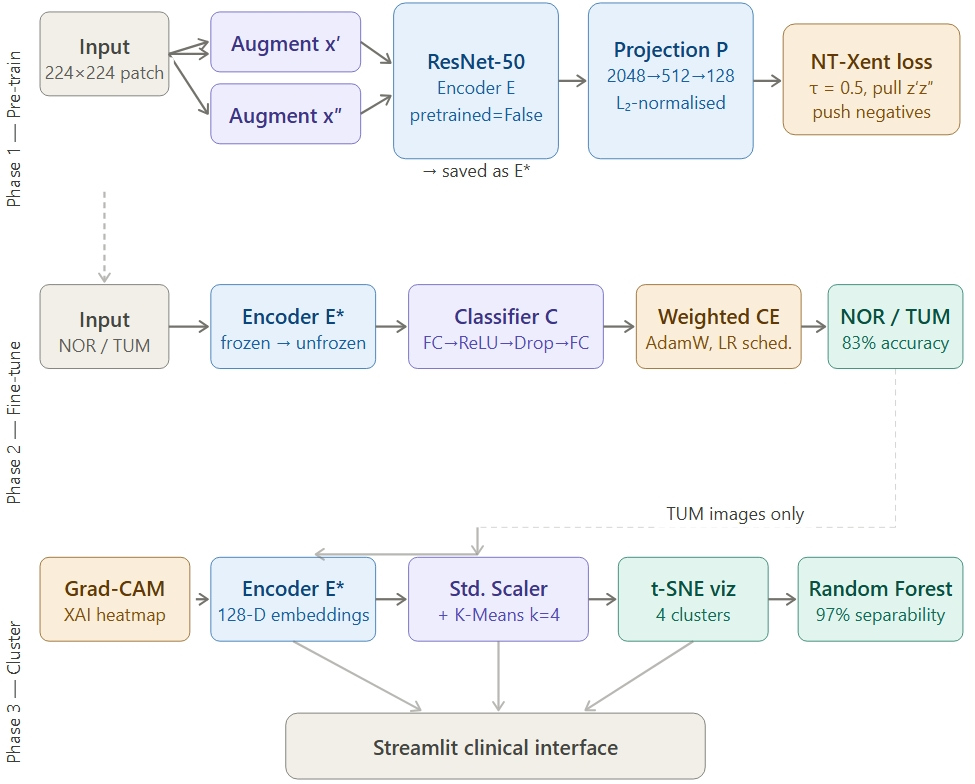}
  \caption{Contrastive learning Technical Framework}
  \label{fig: frame}
\end{figure} 

\subsubsection{Contrastive Learning Encoder}
The contrastive encoder was constructed using ResNet-50 as the backbone architecture. ImageNet pre-training was omitted (pretrained=False) to avoid domain bias because histopathological images exhibit different properties from natural images, and pre-trained features may encode irrelevant textures or colour statistics. The final fully connected layer of ResNet-50 was replaced with an identity mapping to expose the 2048-dimensional feature vector.
A two-layer projection head was appended to map encoder features into a normalised 128-dimensional embedding space: Linear(2048→512) → ReLU → Linear(512→128), followed by L2 normalisation. This projection head is standard in modern CL frameworks such as SimCLR and has been shown to improve the quality of learned representations by separating the task-specific and task-agnostic components of the feature space.
The model was trained using the Normalised Temperature-scaled Cross-Entropy (NT-Xent) loss with temperature T = 0.5. For a batch of N image pairs, NT-Xent treats the two augmented views of the same image as a positive pair and all other 2(N-1) images in the batch as negative samples. The loss encourages alignment of positive pairs while maximising angular separation from negatives in the embedding space. Training was conducted for 20 epochs using the Adam optimiser (lr = $1\times10^{-3}$, batch size = 64).

\subsubsection{Supervised Downstream Classifier}
Following the CL pre-training, the learned encoder was transferred to a supervised binary classification task. The encoder weights were initially frozen, and a lightweight classifier head was appended: Linear(128→128) → ReLU → Dropout(0.3) → Linear(128→2). This architecture allows the classification head to adapt to the labelled training distribution while preserving the general representations learned during contrastive pre-training.
After 3 warm-up epochs with the encoder frozen, the encoder was unfrozen and the entire model fine-tuned using differential learning rates: lr = $1\times10^{-4}$ for encoder parameters and lr = $1\times10^{-3}$ for the classification head, implemented via AdamW with weight decay $1\times10^{-4}$. A ReduceLROnPlateau scheduler (factor = 0.5, patience = 2) was applied based on validation accuracy. Early stopping with patience 5 was used to prevent overfitting. Training was conducted for up to 40 epochs with model check based on best test accuracy.

\subsubsection{Model Evaluation}
The fine-tuned classifier was evaluated on the held-out test set using a comprehensive suite of metrics: overall accuracy, per-class precision, recall, F1-score, ROC-AUC, the Precision-Recall curve, and the confusion matrix. The Dice coefficient was additionally computed per class as: Dice = 2·TP / (2·TP + FP + FN), providing a symmetric measure of predictive overlap. All evaluations were conducted exclusively on the test set, which was not observed during any phase of training.

\subsubsection{Tumor Progression Analysis with Contrastive Clustering}
\begin{figure}[htbp]
  \centering
  \includegraphics[width=3.3 in]{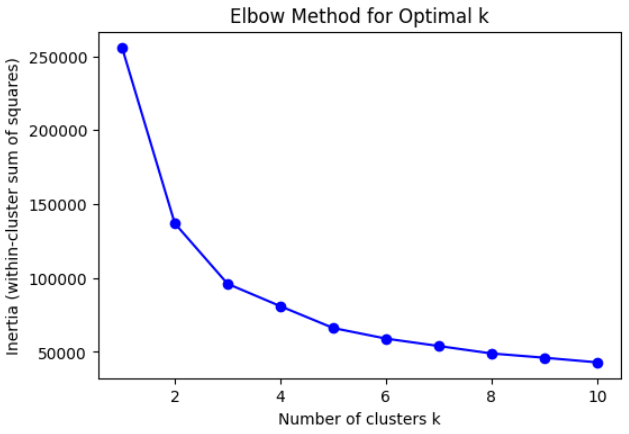}
  \caption{Elbow Method}
  \label{fig: elbow}
\end{figure}

To investigate potential intra-class heterogeneity within tumour samples, which may correspond to distinct progression substages, the trained encoder was used to extract embeddings for all TUM images from both training and test splits.
The resulting embeddings were standardised using StandardScaler before clustering to ensure unit variance across all embedding dimensions. K-Means clustering was then applied to the scaled embeddings. The optimal number of clusters (k = 4) was determined using the Elbow Method as shown in Figure~\ref{fig: elbow} of the plot of within-cluster sum of squares (inertia) against k from 1 to 10.
t-SNE dimensionality reduction was applied to the standardised embeddings to produce a 2D visualisation of the cluster structure. A Random Forest classifier was subsequently trained on the cluster assignments using an 80/20 stratified train-test split to quantify cluster separability. Dice coefficients were computed per cluster to assess intra-cluster consistency.
Essentially, direct clustering will only cluster based on color, orientation and superficial metrics, but a CL encoded embedding can help the clustering algorithm cluster semantically.

\subsubsection{Explainability by Grad-CAM}

For interpretability, Gradient-weighted Class Activation Mapping (Grad-CAM) was integrated into the Streamlit interface. Grad-CAM computes the gradient of the class score with respect to the final convolutional feature maps of the ResNet-50 encoder, producing a spatial heatmap that localises the tissue regions most influential in the classifier's decision. This enables clinicians to visually verify that the model attends to biologically plausible morphological structures  rather than staining artifacts or tissue boundaries, directly addressing the transparency gap identified in the literature review.

\subsubsection{Clinical Interface}

All trained components-the CL encoder, fine-tuned classifier, and clustering pipeline were employed within a Streamlit application to support interactive clinical inference. The interface accepts a histopathological image as input, produces a binary NOR/TUM classification with class probabilities, generates a Grad-CAM heatmap overlaid on the input tissue image, and optionally assigns the image to a tumour progression cluster when classified as TUM. This integrated interface is designed to function as a decision-support tool for pathologists, providing both a diagnostic prediction and an interpretable justification.

\subsection{Algorithm}
Algorithm~\ref{alg:part1} and Algorithm~\ref{alg:part2} present the complete 
procedural flow of the proposed framework across six sequential phases.

\begin{algorithm}[H]
\small
\DontPrintSemicolon
\SetAlgoLined
\caption{CL Framework for Gastric Cancer -- Phases 1--3}
\label{alg:part1}

\KwIn{$\mathcal{D} = \{(\mathbf{x}_i, y_i)\}$, $y_i \in \{\texttt{NOR}, \texttt{TUM}\}$}
\KwOut{Encoder $E^{*}$, classifier $C$}

\BlankLine
\textbf{Phase 1: Data Preparation}\;
Map classes: NOR $\to 0$, TUM $\to 1$\;
Split $\mathcal{D} \to \mathcal{D}_{\text{train}}, \mathcal{D}_{\text{test}}$\;
Apply WeightedRandomSampler $\to$ 6,000 balanced samples/epoch\;

\BlankLine
\textbf{Phase 2: Contrastive Pre-Training}\;
Init ResNet-50 encoder $E$ (pretrained=False); replace fc $\to$ Identity\;
Add projection head: $2048 \to 512 \to 128$, ReLU, $\ell_2$-norm\;
\For{epoch $= 1$ \KwTo $20$}{
  \For{each batch $\{\mathbf{x}_i\}$}{
    $\mathbf{x}_i', \mathbf{x}_i'' \leftarrow$ two augmented views of $\mathbf{x}_i$\;
    $\mathbf{z}_i' = P(E(\mathbf{x}_i'))$,\quad $\mathbf{z}_i'' = P(E(\mathbf{x}_i''))$\;
    $\mathcal{L}_{\text{CL}} = -\log \frac{\exp(\text{sim}(\mathbf{z}',\mathbf{z}'')/\tau)}{\sum_{j\neq i}\exp(\text{sim}(\mathbf{z}',\mathbf{z}_j)/\tau)}$, $\tau{=}0.5$\;
    Update $E, P$ via Adam ($\text{lr}=10^{-3}$)\;
  }
}
Save $E^{*}$\;

\BlankLine
\textbf{Phase 3: Supervised Fine-Tuning}\;
Freeze $E^{*}$; attach $C$: $128 \to 128 \to 2$ (ReLU, Dropout 0.3)\;
\For{epoch $= 1$ \KwTo $3$ \tcp*{warm-up, $C$ only}}{
  Train $C$ via Adam ($\text{lr}=10^{-3}$)\;
}
Unfreeze $E^{*}$; use AdamW: $E^{*}$ lr$=10^{-4}$, $C$ lr$=10^{-3}$\;
\For{epoch $= 4$ \KwTo $40$ \tcp*{full fine-tuning}}{
  \For{each batch $\{(\mathbf{x}_i, y_i)\}$}{
    $\hat{y}_i = C(E^{*}(\mathbf{x}_i))$\;
    $\mathcal{L}_{\text{CE}} = -\sum_i w_{y_i} \log \hat{y}_i$\;
    Update $E^{*}, C$\;
  }
  Step ReduceLROnPlateau; save checkpoint if improved\;
  \lIf{no improvement for 5 epochs}{\textbf{break}}
}

\end{algorithm}

\begin{algorithm}[H]
\small
\DontPrintSemicolon
\SetAlgoLined
\caption{CL Framework for Gastric Cancer -- Phases 4--6}
\label{alg:part2}

\KwIn{Trained $E^{*}$, $C$; test set $\mathcal{D}_{\text{test}}$; TUM images}
\KwOut{Metrics, clustering pipeline $\Psi$, Grad-CAM maps}

\BlankLine
\textbf{Phase 4: Evaluation}\;
Run inference on $\mathcal{D}_{\text{test}}$ (no gradient)\;
Compute: Accuracy, Precision, Recall, F1, ROC-AUC, AP\;
Compute Dice$_c = \frac{2\cdot\text{TP}_c}{2\cdot\text{TP}_c + \text{FP}_c + \text{FN}_c}$ for $c \in \{\texttt{NOR}, \texttt{TUM}\}$\;

\BlankLine
\textbf{Phase 5: Tumour Progression Clustering}\;
Extract $\mathcal{D}_{\text{TUM}}$ from $\mathcal{D}_{\text{train}} \cup \mathcal{D}_{\text{test}}$\;
Compute embeddings $\mathbf{h}_i = E^{*}(\mathbf{x}_i)$; standardise via StandardScaler\;
Find $k^{*}=4$ via Elbow Method (sweep $k=1\ldots10$)\;
Fit KMeans($k{=}4$); assign cluster $c_i$ to each $\mathbf{x}_i$\;
Visualise with t-SNE (2D); train RandomForest (200 trees) on clusters\;
Evaluate separability; serialise $\Psi = \{\text{Scaler, KMeans, RF}\}$\;

\BlankLine
\textbf{Phase 6: Clinical Deployment}\;
Deploy $E^{*}, C, \Psi$ in Streamlit interface\;
\For{each input image $\mathbf{x}_{\text{new}}$}{
  $\hat{y} = C(E^{*}(\mathbf{x}_{\text{new}}))$; output class probabilities\;
  Compute Grad-CAM heatmap $\mathbf{M}$ over final conv layer\;
  \lIf{$\hat{y} = \texttt{TUM}$}{assign progression cluster via $\Psi$}
  Display prediction, confidence, heatmap, cluster label\;
}

\end{algorithm}

Phases 1--4 constitute the core diagnostic pipeline, while Phase 5 extends 
the framework to unsupervised progression analysis. Phase 6 represents the 
clinical translation layer enabling pathologist-facing explainability. 
Figure~\ref{fig: arch} visually depicts the implementation pipeline and 
Figure~\ref{fig: frame} depicts the contrastive learning technical framework.

\section{Results \& Discussion}

The proposed framework was evaluated by two tasks: (i) binary tumour/normal classification using the CL-pretrained ResNet-50 encoder coupled with a supervised downstream classifier, and (ii) unsupervised intra-tumour progression analysis via K-Means clustering of contrastive embeddings. The complete training configuration is summarised in Table ~\ref{tab:conf}. Evaluation was conducted exclusively on the held-out test set, which was not observed during any phase of training or hyperparameter selection.

\begin{table}[h]
\centering
\caption{Training Hyperparameter Configuration for CL Pre-Training and Supervised Fine-Tuning}
\begin{tabular}{|l|c|c|}
\hline
\textbf{Hyperparameter / Setting} & \textbf{CL Pre-Training} & \textbf{Supervised Fine-Tuning} \\ \hline

Backbone & ResNet-50 (no pretrain) & ResNet-50 (CL encoder) \\ \hline

Optimiser & Adam & AdamW \\ \hline

Learning Rate & $1\times10^{-3}$ & Enc: $1\times10^{-4}$ / Head: $1\times10^{-3}$ \\ \hline

Batch Size & 64 & 64 \\ \hline

Epochs & 20 & Up to 40 \\ \hline

Loss Function & NT-Xent ($\tau = 0.5$) & Weighted Cross-Entropy \\ \hline

Embedding Dim. & 128-D ($L_{2}$-normalised) & -- \\ \hline

Dropout & -- & 0.3 \\ \hline

LR Scheduler & -- & ReduceLROnPlateau ($p=2$) \\ \hline

Class Balancing & BalancedCLDataset & WeightedRandomSampler \\ \hline
\end{tabular}
\label{tab:conf}
\end{table}

\begin{figure}[t]
  \centering
  \includegraphics[width=3.3 in]{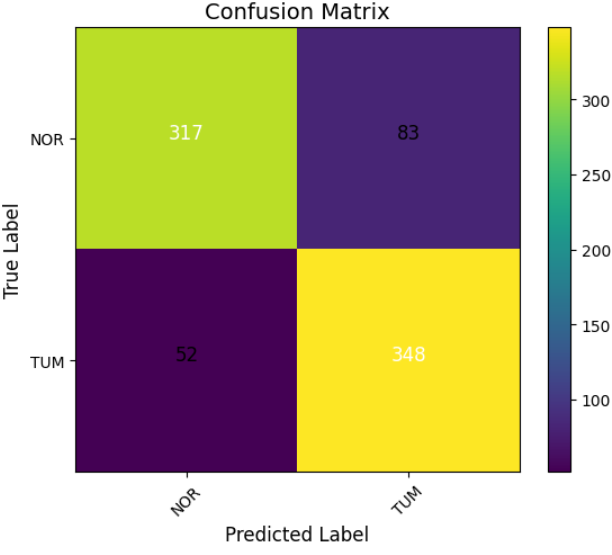}
  \caption{Confusion Matrix (TUM/NOR)}
  \label{fig:conf}
\end{figure} 

\begin{figure}[t]
  \centering
  \includegraphics[width=3.3 in]{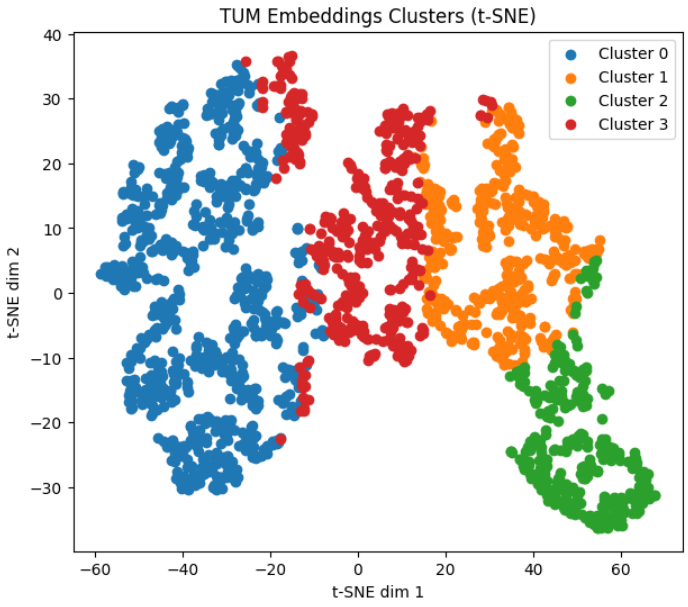}
  \caption{Potential Staging Clusters}
  \label{fig:tsne}
\end{figure} 

\begin{figure}[t]
  \centering
  \includegraphics[width=3.3 in]{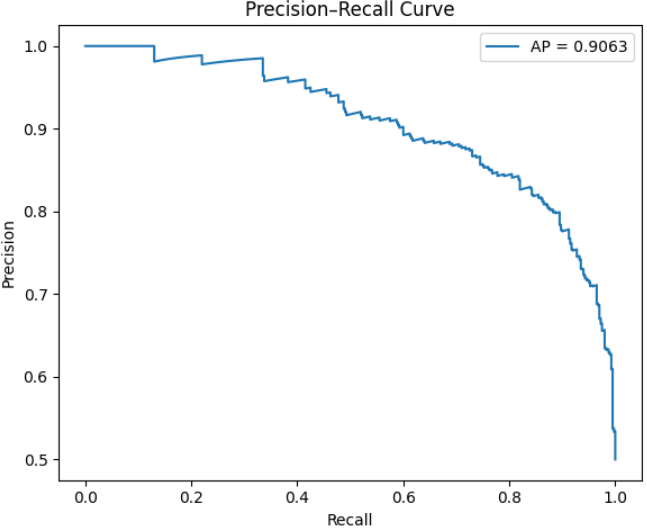}
  \caption{Precision - Recall Curve (TUM/NOR)}
  \label{fig:pr}
\end{figure} 

\begin{figure}[t]
  \centering
  \includegraphics[width=3.3 in]{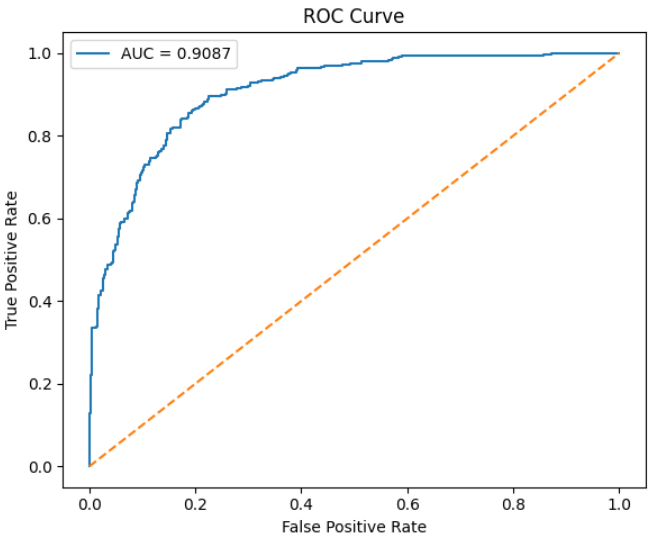}
  \caption{AUC-ROC Curve (TUM/NOR)}
  \label{fig:roc}
\end{figure}

\begin{figure}[t]
  \centering
  \includegraphics[width=3.3 in]{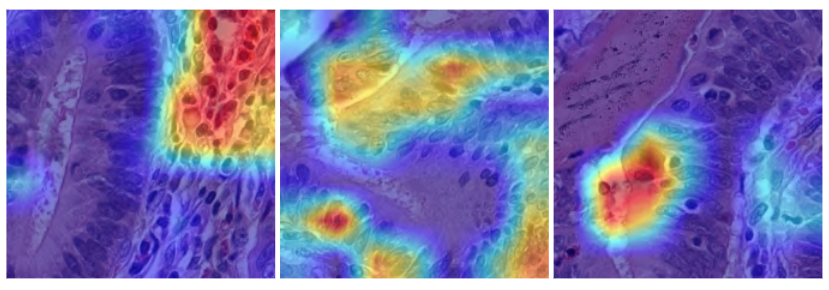}
  \caption{Grad CAM images}
  \label{fig:grad}
\end{figure} 

\begin{figure}[t]
  \centering
  \includegraphics[width=5 in]{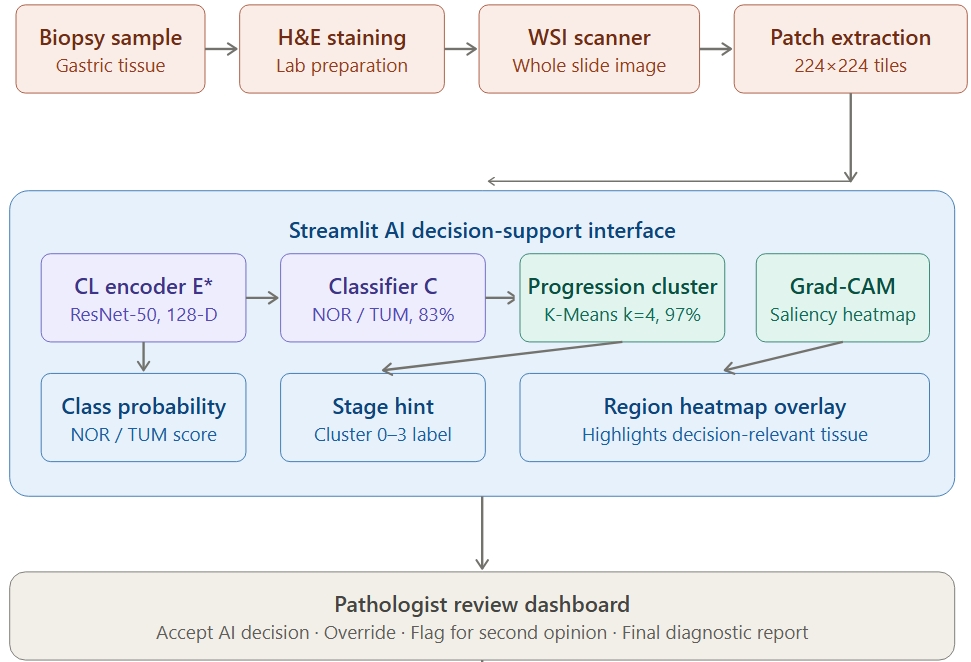}
  \caption{Streamlit Interface Workflow}
  \label{fig:str}
\end{figure} 

\begin{figure*}[htbp]
  \centering
  \includegraphics[width=0.45\textwidth]{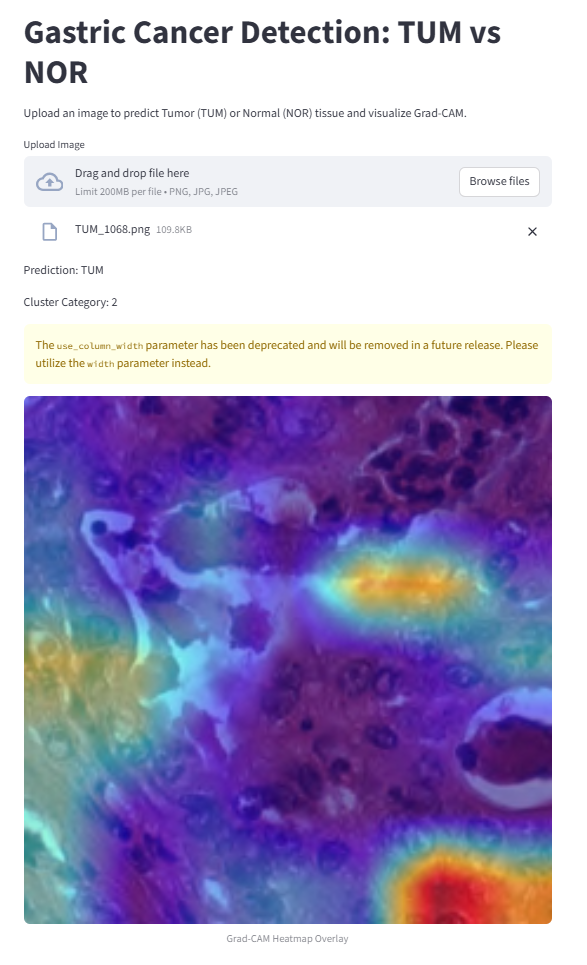}
  \hfill
  \includegraphics[width=0.45\textwidth]{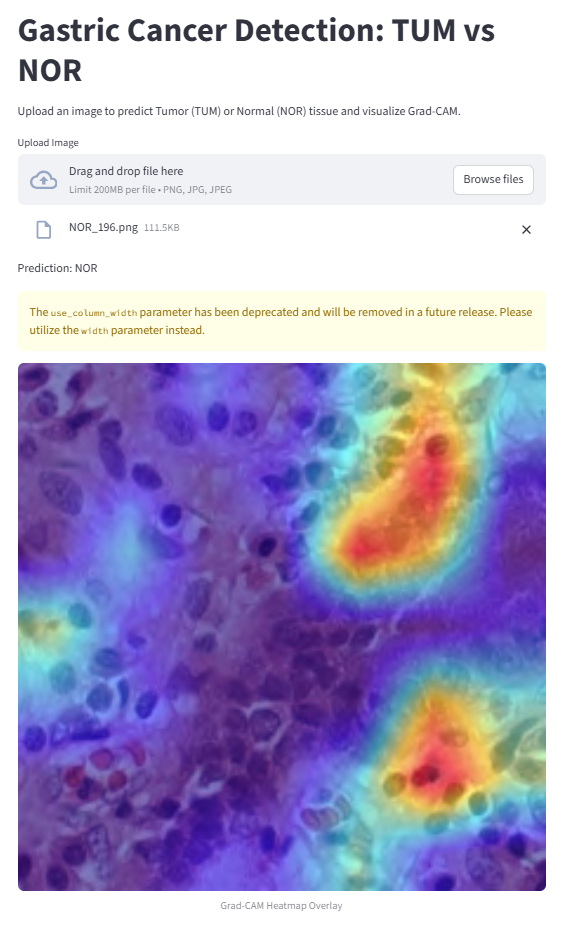}
  \caption{Streamlit Interface}
  \label{fig:int}
\end{figure*}

The fine-tuned classifier achieved a test accuracy of 83.13\% on the binary NOR/TUM classification task. Per-class precision, recall, and F1-score were balanced across both classes, indicating that the model does not exhibit a systematic bias toward either category. The per-class and aggregate metrics are reported in Table ~\ref{tab:met}, and the corresponding confusion matrix is shown in Figure~\ref{fig:conf}.

\begin{table}[h]
\centering
\caption{Classification Performance Metrics}
\begin{tabular}{|l|c|c|c|c|}
\hline
\textbf{Class} & \textbf{Precision} & \textbf{Recall} & \textbf{F1-Score} & \textbf{Dice Coefficient} \\ \hline

NOR (Normal) & 0.8591 & 0.7925 & 0.8244 & 0.8244 \\ \hline

TUM (Tumour) & 0.8074 & 0.8700 & 0.8375 & 0.8375 \\ \hline

Macro Average & 0.8333 & 0.8313 & 0.8310 & 0.8310 \\ \hline

Weighted Average & 0.8333 & 0.8313 & 0.8310 & 0.8310 \\ \hline

\end{tabular}
\label{tab:met}
\end{table}

These tumor images’ embeddings are further clustered into intra-class groups to represent stages or levels of the cancer that is trained on Random Forest Classifier to predict the intra-class stage within tumor images which gives and extra layer of information about disease progressions with limited annotation that can help fasten clinical verifications and annotation processes for better model construction. Figure ~\ref{fig:tsne} presents the t-SNE projection of tumour embeddings into two dimensions, coloured by K-Means cluster assignment. Four spatially coherent 
and well-separated cluster regions are visible. The 
clear geometric separation between clusters - with minimal inter-cluster overlap, confirms that the CL encoder has learned semantically meaningful features that extend beyond superficial pixel-level similarity, encoding morphologically 
distinct tumour substates in the embedding space. This spatial structure is a direct consequence of the NT-Xent objective, which maximises angular distance between dissimilar samples, enabling the K-Means algorithm to discover pathologically coherent groupings rather than arbitrary partitions.

The balanced precision and recall across both classes (NOR: precision=0.8591, recall=0.7925; TUM: precision=0.8074, recall=0.8700) indicates that the model does not exhibit systematic bias toward either category - a clinically important property given the asymmetric consequences of false negatives (missed tumours) and false positives (unnecessary intervention) in cancer diagnosis. The higher recall for the TUM class (0.8700 vs 0.7925 for NOR) is particularly desirable in a screening context, as it prioritises sensitivity to tumour presence. The ROC-AUC of 0.9087 (Figure~\ref{fig:roc}) and Average Precision of 0.9063 (Figure~\ref{fig:pr}) confirm strong discriminative capacity across all 
classification thresholds, indicating that the decision boundary learned from CL embeddings is robust and not merely tuned to the default 0.5 threshold. The proposed self-supervised framework achieves competitive performance while operating with substantially reduced annotation requirements.  Gradient-weighted Class Activation Mapping (Grad-CAM) was applied to the fine-tuned classifier to generate spatial saliency maps highlighting the tissue regions most influential in each prediction as shown in Figure~\ref{fig:grad}. It presents representative Grad-CAM heatmap overlays generated for tumour-classified patches. Warmer colours (red/yellow) indicate regions of highest gradient magnitude with respect to the predicted class, corresponding to areas most influential in the model's decision. Across all three example patches, the activation concentrates on irregular glandular structures and regions of elevated nuclear density, morphological features consistent with the histopathological criteria applied by pathologists in gastric cancer grading. Notably, activation is absent from tissue boundaries, background regions, and areas of uniform staining, indicating that the CL Pre-training has successfully suppressed the encoding of acquisition-related confounds. This spatially coherent explainability output provides pathologists with a mechanism to verify AI predictions against their domain expertise, 
directly addressing the transparency gap identified.
The models built were saved and incorporated into a Streamlit application, wherein users can input images and the model predicts if the image contains a tumor or is normal, and given that it is a tumor, it provides which cluster category it belongs to. This is also accompanied by a Grad-CAM image that depicts which region of the image causes the model to predict a given decision. This integrated output is designed to function as a decision-support tool rather than an autonomous diagnostic system, enabling pathologists to leverage AI-generated insights while retaining clinical authority over final diagnostic decisions as depicted by Figure ~\ref{fig:str} and the interface is depicted by Figure ~\ref{fig:int}. The interface is structured into three parts: (i) an image upload where the clinician submits a histopathological patch; (ii) a prediction displaying the binary NOR/TUM classification; and (iii) an explainability image presenting the Grad-CAM heatmap overlaid on the input image, highlighting decision-relevant tissue regions. When the model predicts TUM, a fourth 
progression grouping activates, displaying the assigned pseudo-stage cluster derived from the K-Means pipeline. The interface is designed to be non-disruptive to clinical workflow, a pathologist can upload a patch extracted from a scanned 
WSI and receive a structured diagnostic output within seconds, with the Grad-CAM 
overlay providing an immediate visual justification that can be cross-referenced against the physical slide.

\section{Conclusion \& Future Scope}
This study presents an integrated framework that enables contrastive learning for gastric cancer diagnosis and addresses the issues of traditional machine learning models. By utilizing ResNet50 based contrastive learning encoder, the model efficiently learns discriminative and generalizable feature representations directly from images and reduces dependency on large-scale datasets. The learned embeddings are then subject to a neural network for binary tumor detection and further clustered to identify pseudo-stages within tumor samples, thus capturing intra-class variations that reflect potential disease progression. A Random Forest Classifier is trained on the cluster labels and confirms high separability and robustness of learned features.

The proposed system achieved an accuracy of 83\% for tumor/normal classification and 97\% in intra-tumor stage categorization that demonstrates strong representational quality and stage level differentiation. Overall, this approach not only helps diagnostic workflows but also bridges the gap between efficiency and clinical utility.
The research can further be improved by validating the intra-tumour clusters identified by K-Means against expert pathological annotations or established clinical staging criteria (e.g., TNM staging). While the high computational separability of the clusters is encouraging, their biological meaning must be confirmed by qualified pathologists before clinical conclusions can be drawn. All experiments were conducted on a single dataset from a single institution, limiting assessment of cross-institutional generalisability which can be expanded. Future research will focus on pathological validation of the identified tumour clusters through collaboration with clinical pathologists. The framework can also be extended to multi-class tissue classification and trained on larger multi-institutional datasets to improve generalisation and robustness. Additionally, integrating whole-slide image analysis with multiple-instance learning and conducting prospective clinical evaluation of the deployment interface will help assess real-world diagnostic utility.

\section{Acknowledgements}
The authors like to express acknowledgment to the developers of the Gastric Cancer Histopathology Images dataset \cite{b26} from Kaggle.

\end{document}